\documentclass[aps,prd,onecolumn,superscriptaddress,showpacs,floatfix,10pt]{revtex4}
\usepackage{graphicx}
\usepackage{amssymb,amsmath}
\usepackage{subfigure}
\usepackage{epsfig}

\begin{document}

\title{Artificially induced positronium oscillations in a two-sheeted spacetime:
consequences on the observed decay processes}

\author{Micha\"{e}l Sarrazin}
\email{michael.sarrazin@fundp.ac.be} \affiliation{Laboratoire de
Physique du Solide, Facult\'es Universitaires Notre-Dame de la Paix,
\\61 rue de Bruxelles, B-5000 Namur, Belgium}

\author{Fabrice Petit}
\email{f.petit@bcrc.be} \affiliation{Belgian Ceramic Research
Centre,\\4 avenue du gouverneur Cornez, B-7000 Mons, Belgium}

\begin{abstract}
Following recent theoretical results, it is suggested that
positronium (Ps) might undergo spontaneous oscillations between two
4D spacetime sheets whenever subjected to constant irrotational
magnetic vector potentials. We show that these oscillations that
would come together with o-Ps/p-Ps oscillations should have
important consequences on Ps decay rates. Experimental setup and
conditions are also suggested for demonstrating in non accelerator
experiments this new invisible decay mode.
\end{abstract}

\pacs{11.27.+d, 11.25.Wx, 11.10.Kk, 36.10.Dr}

\maketitle

\section{Introduction}

In recent years, there has been a renewed interest for Kaluza-Klein
like scenarios suggesting that our usual spacetime could be just a
slice of a larger dimensional manifold \cite{1,2,3,4,5,6,7}. A huge
number of papers have shown that by extending the number of
dimensions, nice explanations of several physical phenomena can be
found. The hierarchy between the electroweak and the Planck scales,
the dark matter origin or the cosmic acceleration can be thus
reasonably addressed in a multidimensional framework. Usually, those
approaches perpetuate the tradition inherited from relativity by
assuming that the whole universe behaves like a smooth continuum
\cite{1,2,3,4,5,6,7}. However, there have been some recent attempts
to develop models where the continuous extra dimensions are
substituted by discrete dimensions \cite{8,9,10,11,12,13,14}. In
those approaches, the extra-dimensions are replaced by a finite
number of points and the whole universe can be seen as made up of a
collection of 4D sheets. Besides keeping the physical richness of
multidimensional spacetimes, such multi-sheeted approaches provide
also a nice framework where the standard model may arise from pure
geometry. For instance, it has been pointed out that ordinary gauge
fields and Higgs field appear spontaneously in some non commutative
models of spacetime \cite{8,9,10}. All these new physical and
mathematical developments are of great interest as they offer new
ways for a better understanding of the building blocks of our
universe.

To date, the physical forces have been accurately measured up to the
weak scale distances. Since no divergence between experimental and
theoretical results has been noted so far, any new physical effect
arising from the existence of extra dimensions is actually severely
constrained. As an illustration of this constraint, standard model
particles are usually expected to be confined on a 4D space manifold
without having any capability of moving throughout the bulk.
Gravitation is the only interaction which is assumed to be able to
connect adjacent spacetimes since otherwise causality and locality
could be violated \cite{4,7}. Nevertheless, this question is still
an open issue that needs to be further clarified. For instance, in
an attempt to circumvent the classical restrictions against particle
motion through the bulk, some works have tried to extend the
hypothetical graviton capability to the case of massive particles as
well \cite{3,5,6}. Along these lines, it was suggested that highly
massive and energetic particles may acquire a non zero five momentum
and escape from a 4-dimensional submanifold to propagate into the
dimensionally extended bulk.

Recently, at the crossroad of brane models and non commutative
two-sheeted spacetimes, present authors have proposed to investigate
the quantum behavior of massive fermions in a $M_{4} \times Z_{2}$
two-sheeted universe. It was suggested that particles of usual
matter might leave our 4D spacetime and reach another distant 4D
spacetime sheet \cite{15,16}. In the proposed framework, the
particle dynamics was studied through two-sheeted extensions of the
classical Dirac, Pauli and Schrodinger equations. Although the usual
particle behavior was recovered in the new framework, the existence
of a tiny geometrical coupling between the two sheets was shown to
provide new interesting physics \cite{15,16}. As the most striking
result, oscillations of massive particles between the two spacetime
sheets were predicted. Accordingly, particles could disappear from
one sheet to reappear in the second one. During oscillations, many
parameters (such as, for instance, baryonic and leptonic numbers or
electric charge) as measured by a $M_{4}$ observer are violated but
they remain globally constant from a $M_{4} \times Z_{2}$ point of
view.

In Refs. \cite{15} and \cite{16}, it was shown that the conditions
for reaching such transfer between both spacetime sheets are far
from trivial. Application of very intense curl-free (irrotational)
magnetic vector potentials was shown to be necessary \footnote{It is
important to recall that if a magnetic field results from a magnetic
vector potential, by contrast, a magnetic vector potential not
necessary induces a magnetic field. In this case, the magnetic
vector potential is called irrotational or curl-free. The particle
disappearance can be obtained with an irrotational magnetic vector
potential only since any magnetic field will tend to inhibit the
effect \cite{15,16}.}. It must be noted that although the result
establishing the oscillations was derived for point like particles,
it was inferred that the equations could be also valid for more
complex particles like protons or neutrons for instance, the
essential point being that the particle exhibits a non zero magnetic
moment \cite{15,16}.

To address the issue of whether quantum motions between two
spacetime sheets are indeed possible, the present paper discusses
the possibility of a controlled production of these oscillations in
the case of positronium (Ps). This choice was dictated for two
reasons. Firstly, positronium is at the heart of many experimental
researches \cite{17,18,19,20,21} and secondly, there is already a
wealth of papers suggesting that positronium could possibly exhibit
an invisible decay \cite{19,20,21}. Since each model of invisible
positronium decay possesses its own specific signature, it is
expected that the experimental confirmation of such decay could give
insights on the form and size of the extra dimension(s)
\cite{19,20,21}. The purpose of this paper is then to clarify how
such a Ps invisible decay would looks like in a $M_{4} \times Z_{2}$
two-sheeted spacetime as described in Refs. \cite{15} and \cite{16}.

\section{Mathematical framework}

In Refs. \cite{15} and \cite{16}, a model describing the quantum
dynamics of matter particles in a two-sheeted spacetime was
introduced. It formally corresponds to the product of a four
continuous manifold times a discrete two points space, i.e. $X =
M_{4} \times Z_{2}$. The model was built following two different
approaches which are now briefly summarized.
\\The former approach was mainly based on the work of
Connes \cite{8,9}, Viet and Wali \cite{11,12} and relies on a
non-commutative definition of the exterior derivative acting on the
product manifold. Due to the specific geometrical structure on the
discrete space, this operator is given by :

\begin{equation}
D_\mu =\left( {{
\begin{array}{cc}
{\partial _\mu } & 0 \\
0 & {\partial _\mu }
\end{array}
}}\right) ,\text{ }\mu =0,1,2,3\text{ and\ }D_5=\left( {{
\begin{array}{cc}
0 & g \\
{-g} & 0
\end{array}
}}\right)   \label{1}
\end{equation}
Where the parameter $g$ has the dimension of mass (in $\hbar = c =
1$ units) and acts as a finite difference operator along the
discrete dimension. In the model discussed in Ref. \cite{15} and
contrary to previous works \cite{8}, $g$ was considered as a
constant geometrical field (not the Higgs field). As a consequence a
mass term was introduced and a two-sheeted Dirac equation was
derived :

\begin{equation}
\left( {iD\!\!\!/ -M}\right) \Psi =\left( {i\Gamma ^ND_N-M}\right)
\Psi =\left( {{
\begin{array}{cc}
{i\gamma ^\mu \partial _\mu -m} & {ig\gamma ^5} \\
{ig\gamma ^5} & {i\gamma ^\mu \partial _\mu -m}
\end{array}
}}\right) \left( {{
\begin{array}{c}
{\psi _{+}} \\
{\psi _{-}}
\end{array}
}}\right) =0  \label{2}
\end{equation}

It can be noticed that by virtue of the two-sheeted structure of
spacetime, the wave function $\psi$ of the fermion comprises two
components located on different 4D sheets. Therefore in this model,
any massive particles is in fact a five dimensional entity. Although
this $M_{4} \times Z_{2}$ delocalization can be seen as a major
drawback of the model, it was demonstrated that environmental
interactions force the particle localization in one or the other
sheet and act as a natural dimensional reduction mechanism
\cite{15,16}.

Obviously, the equation (2) can also be derived from the $M_{4}
\times Z_{2}$ lagrangian \textbf{\textit{L}} defined as
\begin{equation}
\textbf{\textit{L}}=\bar{\Psi} \left( {i\not{D}-M}\right) \Psi
\label{3}
\end{equation}
Where $\bar{\Psi}=\left( {\bar{\psi}_{+},\bar{\psi}_{-}}\right) $ is
the two-sheeted adjoint spinor of $\Psi $ and with $\bar{\psi}_{+}$ and $\bar{\psi}%
_{-}$ the adjoint spinors respectively of $\psi _{+}$ and $\psi
_{-}$.

Such a lagrangian can be also written into the following expended
form
\begin{equation}
\textbf{\textit{L}}=\bar{\psi}_{+} \left( {i\not{\partial}-m}\right)
\psi _{+} + \bar{\psi}_{-} \left( {i\not{\partial}-m}\right) \psi
_{-} + ig\bar{\psi}_{+}\gamma ^5\psi _{-} + ig\bar{\psi}_{-}\gamma
^5\psi _{+}\label{4}
\end{equation}

At first sight, the doubling of the wave function can be seen as a
reminiscence of the hidden-sector concept. While it is true that
hidden sector models and present approach share several common
points, it is equally true that they differ by the number of
spacetime sheets they consider. For instance, the so-called Mirror
matter approach, considers only one 4D manifold and justifies for
the left/right parity by introducing new internal degrees of freedom
to particles (see for instance Ref. \cite{22}). In the present work,
it can be noted that the number of particle families remains
unchanged but the particles have now access granted to two distinct
4D spacetime sheets.

To be consistent with the new framework the usual U(1) gauge field
must be substituted by an extended U(1)$\bigotimes$U(1) gauge
relevant for the discrete $Z_{2}$ structure of the universe. In
addition, each sheet possesses its own current and charge density
distribution as source of the electromagnetic field. The most
general form for the new gauge (to be incorporated within the
two-sheeted Dirac equation such that $D\!\!\!/ \rightarrow D\!\!\!/
+ A\!\!\!/$) is then defined by (see also Refs. \cite{13} and
\cite{14} where such a gauge was also considered)

\begin{equation}
A\!\!\!/ =\left( {{
\begin{array}{cc}
{iq\gamma^{\mu}A_{+,\mu}} & {\gamma^{5} \chi} \\
{\gamma^{5} \chi^{\dagger}} & {iq\gamma^{\mu}A_{-,\mu}}
\end{array}
}}\right)   \label{5}
\end{equation}

On the two sheets live then the distinct $\textbf{A}_{+}$ and
$\textbf{A}_{-}$ fields. In the present model, the component $\chi$
cannot be associated to the usual Higgs field encountered in GSW
model. An obvious consequence of this off-diagonal term $\chi$ is to
couple the photons fields of the two sheets. Since this term
introduces major complications unless to be weak enough compared to
$\textbf{A}_{\pm}$ (most notably it leads to unusual transformation
laws of the electromagnetic field which are difficult to reconcile
with observations) it is logical to set $\chi=0$. The
electromagnetic fields of both sheets are then completely decoupled
and each sheet is endowed with its own electromagnetic structure.
Note that the another consequence of having $\chi\neq 0$ is to
couple each charged particle with the electromagnetic fields of both
sheets, irrespective of the localization of this particle in the
discrete space. For instance a particle of charge $e$ localized in
the "$+$" sheet would have been sensitive to the electromagnetic
field of the "$-$" sheet with an effective charge $\varepsilon e$
($\varepsilon < 1$). This kind of exotic interactions has been
considered previously in literature within the framework of mirror
matter paradigm and is not covered by the present paper
\cite{23,24}. Another noticeable consequence of considering $\chi=0$
is that the photons fields are now totally trapped in their original
sheets : photons belonging to a given sheet are not able to go into
the other sheet (and as a consequence, structures belonging to a
given sheet are invisible from the perspective of an observer
located in the other sheet). The classical gauge field
transformation which reads

\begin{equation}
A_\mu ^{\prime }=A_\mu +\partial _\mu \Lambda \label{6}
\end{equation}

can now be easily extended in the $Z_{2}$ five dimensional framework
using Eq. (5). To be consistent with the above hypothesis (notably
$\chi=0$), it can be shown that the gauge transformation is
degenerated and reduced to a single $e^{iq\Lambda}$ which must be
applied to both sheets simultaneously \cite{16}. By setting $\chi=0$
and by considering the same gauge transformation in the two sheets
we can get photons fields $\textbf{A}_{\pm}$ which behave
independently from each other and in accordance with observations
\cite{15,16}.

After introducing the gauge field into the $Z_{2}$ Dirac equation
and taking the non relativistic limit (following the standard
procedure), a two-sheeted Pauli like equation can be derived
\begin{equation}
i\hbar \frac \partial {\partial t}\left| \Psi \right\rangle =\mathbf{H}%
\left| \Psi \right\rangle   \label{7}
\end{equation}
with $\left| \Psi \right\rangle =\left(
\begin{array}{c}
\left| \psi _{+}\right\rangle  \\
\left| \psi _{-}\right\rangle
\end{array}
\right) $, where $\left| \psi _{+}\right\rangle $ and $\left| \psi
_{-}\right\rangle $ correspond to the wave functions in the $(+)$
and $(-)$ sheets respectively. The Hamiltonian $\mathbf{H}$ involves
different contributions \cite{15,16}, listed below (using now
natural units)

\begin{equation}
\mathbf{H}_k=-\frac{\hbar ^2}{2m}\left[
\begin{array}{cc}
\left( \mathbf{\nabla }-i\frac q\hbar \mathbf{A}_{+}\right) ^2 & 0 \\
0 & \left( \mathbf{\nabla }-i\frac q\hbar \mathbf{A}_{-}\right) ^2
\end{array}
\right]   \label{8}
\end{equation}
\begin{equation}
\mathbf{H}_m=-g_s\mu \frac \hbar 2\left[
\begin{array}{cc}
\mathbf{\sigma \cdot B}_{+} & 0 \\
0 & \mathbf{\sigma \cdot B}_{-}
\end{array}
\right]   \label{9}
\end{equation}
\begin{equation}
\mathbf{H}_p=\left[
\begin{array}{cc}
V_{+} & 0 \\
0 & V_{-}
\end{array}
\right]   \label{10}
\end{equation}
\begin{equation}
\mathbf{H}_{cm}=ig\gamma g_s\mu \frac \hbar 2\left[
\begin{array}{cc}
0 & \mathbf{\sigma \cdot }\left\{ \mathbf{A}_{+}-\mathbf{A}_{-}\right\}  \\
-\mathbf{\sigma \cdot }\left\{ \mathbf{A}_{+}-\mathbf{A}_{-}\right\}
& 0
\end{array}
\right]   \label{11}
\end{equation}
\begin{equation}
\mathbf{H}_c=\frac{g^2\hbar ^2}m\left[
\begin{array}{cc}
1 & 0 \\
0 & 1
\end{array}
\right]   \label{12}
\end{equation}
where the first three terms, i.e. $\mathbf{H}_k$, $\mathbf{H}_m$ and $%
\mathbf{H}_p$, are reminiscent of the terms found in the classical
Pauli equation in presence of an electromagnetic field.
$\mathbf{A}_{+}$ and $\mathbf{A}_{-}$ denote the magnetic potential
vectors in the sheet (+) and (-) respectively. The same convention
is used for the magnetic fields. $\mathbf{H}_k$ relates to the
kinetic part and includes the vector potential, $\mathbf{H}_m$ is
the coupling term between the magnetic field and the magnetic moment
of the particle $g_s\mu $ where $g_s$ is the gyromagnetic factor and
$\mathbf{H}_p$ the coulomb term. In addition to these ``classical''
terms, the hamiltonian comprises also a specific term involving an
"electromagnetic coupling" between the two sheets. Note that this
coupling $\mathbf{H}_{cm}$ arises through the magnetic vector
potential and the magnetic moment only. For an electron, it can be
shown that $\gamma =1$ though it is clear that it could differ
slightly from unity due to QED corrections in the case of composite
particles like proton or neutron \cite{16}. It can be noted that
$\mathbf{H}_{c}$ is a simple constant term which can be obviously
eliminated through an appropriate rescaling of the energy. The exact
physical meaning of this term shall not be discussed here. In the
model, $g$ is the coupling constant between the two sheets.
$g^{-1}$ is simply the distance between the two sheets \cite{15,16}.\\
The second approach developed in Ref. \cite{16} starts from the
usual covariant Dirac equation in 5D. Assuming a discrete structure
of the discrete space with two four dimensional submanifolds
requires to substitute the extra dimensional derivative by a finite
difference counterpart defined as
\begin{equation}
(\partial _5\psi )_{\pm }=\pm g(\psi _{+}-\psi _{-})  \label{13}
\end{equation}
Then as previously for the non-commutative approach, the Dirac
equation breaks down into a set of two coupled differential
equations similar to Eq. (2) \cite{16}. After introducing
electromagnetic fields into the model and taking the non
relativistic limit, it was shown that the two-sheeted Pauli's
equation takes the same form than Eq. (7) to (12) except for the
$\mathbf{H}_c$ term
\begin{equation} \mathbf{H}_c=\frac{g^2\hbar ^2}m\left[
\begin{array}{cc}
1 & -1 \\
-1 & 1
\end{array}
\right]   \label{14}
\end{equation}
By contrast to the previous $\mathbf{H}_c$ term, this new one cannot
be made vanishing through a simple rescaling of the energy scale.
The off diagonal part of $\mathbf{H}_c$ was demonstrated to be
responsible of particle oscillations between the two spacetime
sheets, even in the case of a free particle \cite{16}. Nevertheless,
as explained in Ref.\cite{16}, $\mathbf{H}_c$ can be neglected, in a
first approximation, in comparison with $\mathbf{H}_{cm}$ since it
is proportional to $g^2$, an obvious tiny value in order for the
model to be consistent with known experimental results. As a
consequence the classical finite difference \cite{16} and the non
commutative \cite{15} approaches of the two-sheeted spacetime
predict almost the same physics. The only difference arises from the
presence of this tiny coupling term $\mathbf{H}_c$ in the finite
difference approach. In this paper we will concentrate mainly on the
coupling term $\mathbf{H}_{cm}$ proportional to $g$ which is
expected to impart most of the new physics contained in both models
\cite{15,16}.

\section{Invisible Positronium decay in a two-sheeted spacetime}

To illustrate how the two-sheeted geometrical structure of spacetime
governs the quantum behavior of particles, let us examine its
effects on the Ps decay processes. At the fundamental state $1S$
(about $-6.8$ eV), Ps exists in two forms. The first form $1^3S_1$
called ortho-positronium (o-Ps), corresponds to parallel orientation
of electron and positon spins. The second form $1^1S_0$ or
para-positronium (p-Ps), corresponds to antiparallel spins states.
O-Ps decay time is about $1,4\cdot 10^{-7}$ s with an emission of
three photons whereas p-Ps decay time is about $1,25\cdot 10^{-10}$
s with an emission of two photons. Besides, the hyperfine structure
interval of positronium ($1^3S_1\rightarrow 1^1S_0$ ) is about
$8.41\cdot 10^{-4}$ eV (i.e. $\sim 203.39$ GHz) \cite{17,18}.
Following the standard description of Ps, the o-Ps wave function can
be expressed as \cite{17,25}
\begin{equation}
\left| \psi _1\right\rangle =\left| \varphi _e\left( \uparrow
\right) \right\rangle \otimes \left| \varphi _p\left( \uparrow
\right) \right\rangle \label{15}
\end{equation}
\begin{equation}
\left| \psi _{-1}\right\rangle =\left| \varphi _e\left( \downarrow
\right) \right\rangle \otimes \left| \varphi _p\left( \downarrow
\right) \right\rangle  \label{16}
\end{equation}
\begin{equation}
\left| \psi _0\right\rangle =(1/\sqrt{2})\left\{ \left| \varphi
_e\left( \uparrow \right) \right\rangle \otimes \left| \varphi
_p\left( \downarrow \right) \right\rangle +\left| \varphi _e\left(
\downarrow \right) \right\rangle \otimes \left| \varphi _p\left(
\uparrow \right) \right\rangle \right\}  \label{17}
\end{equation}
and similarly for the p-Ps wave function, we have
\begin{equation}
\left| \psi \right\rangle =(1/\sqrt{2})\left\{ \left| \varphi
_e\left( \uparrow \right) \right\rangle \otimes \left| \varphi
_p\left( \downarrow \right) \right\rangle -\left| \varphi _e\left(
\downarrow \right) \right\rangle \otimes \left| \varphi _p\left(
\uparrow \right) \right\rangle \right\}  \label{18}
\end{equation}
where $\left| \varphi _e\left( \updownarrow \right) \right\rangle $ and $%
\left| \varphi _p\left( \updownarrow \right) \right\rangle $ relate
to the wave functions of the electron and positon respectively both
taking into account the spin state.

We now consider the influence of a constant irrotational magnetic
vector potential $\mathbf{A}$ located in our own spacetime sheet
(arbitrarily taken to be the (+) sheet). For reasons explained
before, let us also neglect all other terms in the Hamiltonian
except the coupling

\begin{equation}
\mathbf{W}=ig\gamma g_s\mu \frac \hbar 2\left[
\begin{array}{cc}
0 & \left[ \mathbf{\sigma }_e\mathbf{-\sigma }_p\right] \mathbf{\cdot A} \\
-\left[ \mathbf{\sigma }_e-\mathbf{\sigma }_p\right] \mathbf{\cdot
A} & 0
\end{array}
\right]  \label{19}
\end{equation}

Note that this term is simply the sum of two $\mathbf{H}_{cm}$
hamiltonian operators corresponding to the contributions of the
electron and the positon. The minus sign arises from the opposite
magnetic moment of these particles.

Choosing $\mathbf{A=}A\mathbf{e}_z$ for instance and noticing that
\begin{equation}
\left[ \sigma _{z,e}-\sigma _{z,p}\right] \left| \psi
_0\right\rangle =2\left| \psi \right\rangle  \label{20}
\end{equation}
\begin{equation}
\left[ \sigma _{z,e}-\sigma _{z,p}\right] \left| \psi \right\rangle
=2\left| \psi _0\right\rangle  \label{21}
\end{equation}
and
\begin{equation}
\left[ \sigma _{z,e}-\sigma _{z,p}\right] \left| \psi _{\pm
1}\right\rangle =0  \label{22}
\end{equation}
one can see that the only states connected through the term (19) are
of the form
\begin{equation}
\left| \Psi _0\right\rangle =\left[
\begin{array}{c}
\left| \psi _0\right\rangle \\
0
\end{array}
\right]  \label{23}
\end{equation}
and
\begin{equation}
\left| \Psi \right\rangle =\left[
\begin{array}{c}
0 \\
\left| \psi \right\rangle
\end{array}
\right]  \label{24}
\end{equation}
such that the corresponding matrix terms of $\mathbf{W}$ take the
form
\begin{equation}
\left\langle \Psi \right| \mathbf{W}\left| \Psi _0\right\rangle
=-ig\gamma g_s\mu \hbar A  \label{25}
\end{equation}
\begin{equation}
\left\langle \Psi _0\right| \mathbf{W}\left| \Psi \right\rangle
=ig\gamma g_s\mu \hbar A  \label{26}
\end{equation}
We now look for a solution of the form
\begin{equation}
\left| \Phi (t)\right\rangle =a_0(t)e^{-iE_0t/\hbar }\left| \Psi
_0\right\rangle +a_p(t)e^{-iE_pt/\hbar }\left| \Psi \right\rangle
\label{27}
\end{equation}
with $\left| \Phi (t=0)\right\rangle =\left| \Psi _0\right\rangle $.
Putting Eq. (27) into the Schrodinger equation leads to the
following system of coupled differential equations
\begin{equation}
\frac d{dt}a_0=\kappa a_pe^{i\omega _0t}-\Gamma _oa_0  \label{28}
\end{equation}
\begin{equation}
\frac d{dt}a_p=-\kappa a_0e^{-i\omega _0t}-\Gamma _pa_p \label{29}
\end{equation}
with $\kappa =g\gamma g_s\mu A$.

Note that the decay constants $\Gamma _o$ and $\Gamma _p$ have been
conveniently introduced in the equations in agreement with the known
lifetime of the Ps states, i.e. $\Gamma _o\sim 3.57\cdot 10^6$ s$^{-1}$and $%
\Gamma _p\sim 4\cdot 10^9$ s$^{-1}$. In addition we set $\omega
_0=(E_0-E_p)/\hbar \sim 1.28\cdot 10^{12}$ s$^{-1}$ corresponding to
the hyperfine structure interval of Ps. The solution of the above
system is straightforward
\begin{equation}
a_0 =\frac 1{4\sigma }e^{-(1/2)(\Gamma _o+\Gamma _p)t}\left[ \left(
(\Gamma _o-\Gamma _p+i\omega _0)+2\sigma \right) e^{-\sigma
t}\right. \left. -\left( (\Gamma _o-\Gamma _p+i\omega _0)-2\sigma
\right) e^{\sigma t}\right] e^{(i/2)\omega _0t} \label{30}
\end{equation}
and
\begin{equation}
a_p=\frac \kappa {2\sigma }e^{-(1/2)(\Gamma _o+\Gamma _p)t}\left[
e^{-\sigma t}-e^{\sigma t}\right] e^{-(i/2)\omega _0t}  \label{31}
\end{equation}
with
\begin{equation}
\sigma =\frac 12\sqrt{(\Gamma _o-\Gamma _p+i\omega _0)^2-4\kappa ^2}
\label{32}
\end{equation}
A first interesting case to be considered is the one where $\Gamma _o$ and $%
\Gamma _p$ are both equal to zero (the natural decay processes are
not taken into account). Then, it can be shown that
\begin{equation}
\left| a_p\right| ^2=\frac{\kappa ^2}{\rho ^2}\sin ^2gt \label{33}
\end{equation}
and $\left| a_0\right| ^2=1-\left| a_p\right| ^2$, with $\rho =(1/2)\sqrt{%
\omega _0{}^2+4\kappa ^2}$.

\begin{figure}[tpb]
\centerline{\psfig{file=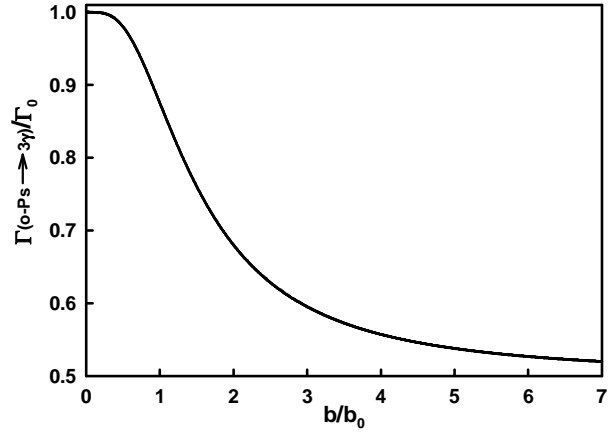,width=8cm}} \vspace*{8pt}
\caption{Theoretical decay rate of o-Ps under the influence of an
irrotational constant magnetic vector potential. \label{f1}}
\end{figure}

This term simply relates to the probability of finding the Ps in the
second spacetime sheet at time $t$. Clearly, Ps oscillates between
the two sheets. The average probability of finding the Ps in the
second sheet is simply given by $\kappa ^2/2\rho ^2$. Since Ps was
assumed to be in a o-Ps state in the first sheet, it is clear that
the larger $\kappa ^2/2\rho ^2$ is, the larger is the probability
for an invisible decay with a p-Ps state in the second sheet. To get
an experimental confirmation of this result, it will be interesting
to achieve experimental conditions such that $\kappa $ $\gg \omega
_0/2$. Let us now return to the more general case including natural
decay modes also. We are now going to calculate the whole decay rate
$\Gamma $. The probabilities for each Ps state, $P_o=\left|
a_0\right| ^2$ and $P_p=\left| a_p\right| ^2$ allow to define the Ps
decay probability $P=1-P_o-P_p$. The related distribution function
for the decay probability is $f(t)=(d/dt)P$ such that the Ps
lifetime is then given by $\tau =\int_0^\infty tf(t)dt$ and the
decay rate by $\Gamma =(1/2)\tau ^{-1}$. Using Eq. (30) and Eq.
(31), $\Gamma $ can be easily derived
\begin{equation}
\Gamma =\frac{(\Gamma _o+\Gamma _p)^2(\kappa ^2+\Gamma _o\Gamma
_p)+\Gamma _o\Gamma _p\omega _0^2}{2\kappa ^2(\Gamma _o+\Gamma
_p)+\Gamma _p((\Gamma _o+\Gamma _p)^2+\omega _0^2)}  \label{34}
\end{equation}
Reminding that $\omega _0\gg \Gamma _p>\Gamma _o$, the above
solution can be approximated with a very good accuracy by
\begin{equation}
\Gamma =\Gamma _o(1-(1/2)(1+\eta ^2)^{-2})+(1/2)\Gamma _p(1+\eta
^2)^{-1} \label{35}
\end{equation}
where $\eta =\omega _0/(\kappa \sqrt{2})$. The following
substitution $\eta =b_0/b$ with $b=gA$ and $b_0=\omega
_0m_e/(e\sqrt{2})$, appears to be convenient for displaying the
solution since the actual value of $g$ is presently unknown. Note
that $b$ has the dimension of a magnetic field like $b_0$. In the
present case $b_0\sim 5.14$ T.
From Eq. (35), the interpretation of
the global decay rate can be made quite easily. The first term
corresponds to the decay rate of the o-Ps in our spacetime sheet,
i.e.
\begin{equation}
\Gamma (\text{o-Ps}\rightarrow 3\gamma )=\Gamma _o(1-(1/2)(1+\eta
^2)^{-2}) \label{36}
\end{equation}

The second term relates to the decay rate of the p-Ps after its
conversion from o-Ps to p-Ps right after its ``jump'' in the second
sheet (and seen from the perspective of an observer living in the
second spacetime sheet)
\begin{equation}
\Gamma (\text{o-Ps}\rightarrow \text{p-Ps}^{\prime }\rightarrow
2\gamma ^{\prime })=(1/2)\Gamma _p(1+\eta ^2)^{-1}  \label{37}
\end{equation}

From an experimental point of view, our results suggest that an o-Ps
population in a constant irrotational vector potential should
present an abnormal decay rate according to Eq. (36) (see Fig.1). A
noticeable point is that one should observe a decreasing photon
production rate as the magnetic potential increases, such a behavior
being related to the p-Ps invisible decay in the second sheet.

It must be pointed out that these results remain unchanged if the
initial o-Ps state is substituted by a p-Ps state (except for a
permutation of the $\Gamma _o$ and $\Gamma _p$ terms in expressions
(34) and (35) respectively).\\
In previous works \cite{15,16}, it was mentioned that a constant
irrotational potential \textbf{A} could be achieved in a hollow
conducting cylinder with a constant current flow $I$ inside. For
such a configuration, we get typically $A\sim \mu_{0}I$
\footnote{Note that an equivalent although perhaps more convenient
geometry than an hollow cylinder can be used. It consists of a
complex toroidal coil and was described in Ref. \cite{26}.}. So, for
a specific decay rate $\Gamma$, we can expect to measure the
variation $\Delta\Gamma$ of the decay as $A$, i.e. $I$ varies. For
the proposed experimental setup, the sensitivity of the apparatus
$\Delta\Gamma /\Gamma$ would thus be simply related to the current
value circulating in the cylinder shell. The figure 2 shows the
value of the coupling constant we can thus expect to measure for
different achievable intensities \footnote{Note that the
reintroduction of the offdiagonal terms of $\mathbf{H}_c$ also leads
to an invisible decay of Ps in the second spacetime sheet.
Nevertheless a careful analysis shows that $\Gamma$ weakly decreases
according to $\Delta \Gamma /\Gamma \sim (1/2)(\chi ^2/\Gamma ^2)$
where $\chi =2\hbar g^2/m_e$. Since this term is proportional to the
square of the coupling constant, it can be neglected. For instance,
even with $g\sim 10^3$ m$^{-1}$ (i.e. a distance about $1$ mm
between sheets) one obtains relative variations of the decay rate of
nearly $2\cdot 10^{-9}$ and $2\cdot 10^{-15}$ for o-Ps and p-Ps
respectively. In that case, Ps freely oscillates between sheets
without p-Ps/o-Ps conversion. It is a matter of evidence that such a
low contribution to the decay rate is still far from the present
experimental accuracies \cite{17,18,21}.}. Reciprocally, since the
value of the coupling constant is presently unknown, the current
required to produce vector potentials allowing particle oscillations
can be hardly estimated. However, it is very likely that the
required intensities are above the present technology capabilities.

\begin{figure}[tpb]
\centerline{\psfig{file=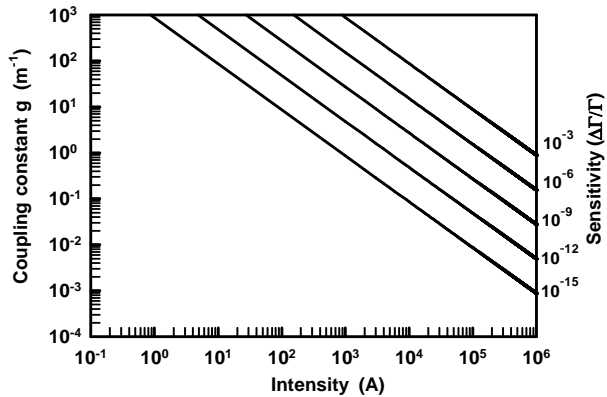,width=8cm}} \vspace*{8pt}
\caption{Measurable coupling constant vs current intensity for
various apparatus sensitivities. \label{f2}}
\end{figure}

In Ref. \cite{23} dealing with mirror matter, a mixing between o-Ps
and mirror o-Ps was also suggested. However the model treated in
this paper \cite{23} is different from that considered in this work
in several aspects. First, this paper \cite{23} does not localize
mirror matter on a distinct spacetime sheet. Secondly, it considers
that the mixing between both worlds occurs through a photon/mirror
photon kinetic mixing (gravitation also is assumed to mediate
interactions between matter and mirror matter but it is not relevant
in the present discussion). Typically, this mixing is then given by
the Lagrangian \cite{23}
\begin{equation}
L = -\varepsilon F_{\mu\nu}F^{'\mu\nu} \label{38}
\end{equation}
with $\varepsilon$ the coupling strength. Considering matter/mirror
matter quantum electrodynamics with this coupling term shows that
oscillations of o-Ps into mirror o-Ps effectively occur in a vacuum
experiment. The result is an apparent invisible decay similar to
that proposed in the present paper. Nevertheless, it must be
stressed that an obvious advantage of the two-sheeted structure
discussed presently is to demonstrate the existence of oscillations
without recourse to a photon/mirror photon kinetic mixing (thus
keeping safe the usual electromagnetic laws). Most important, our
model suggests a way to artificially modify the decay rate of
positronium. At present time, the possible experimental signals for
the existence of extra-dimensions are limited. They rely mainly on
the observation of deviations from the inverse square law of gravity
or spectrum determination of Kaluza-Klein tower states. The
two-sheeted spacetime paradigm suggests a new possibility involving
usual matter particles oscillations between adjacent 4D sheets. An
artificial control of the positronium decay rate by using a device
similar to that proposed in this paper should clearly be
investigated. If an abnormal decay rate could indeed be noticed by
applying an irrotational magnetic potential onto Ps, then it could
reveal the existence of another spacetime sheet and confirms that
our spacetime is just a sheet embedded in a more complex manifold.

\section{Conclusion}

In this paper, we have sketched a possible experimental consequence
of the two-sheeted spacetime introduced in Refs. \cite{15} and
\cite{16}. Ps decay was just considered here as an illustration of
the model but more complex situations can be easily envisaged. It
has been demonstrated that invisible decay of Ps should occur in
presence of a constant irrotational magnetic vector potential. As
described, the invisible decay occurs through particle oscillations
between two distinct spacetime sheets. A simple experimental setup
designed to produce such oscillations has been suggested. The
predictions of the model could be relevant for research teams aiming
at demonstrating abnormal Ps behavior not accommodated by the
standard model.

\end{document}